\documentclass[preprint,aps,tightenlines,floatfix,showpacs,byrevtex,nofootinbib]{revtex4}
\usepackage{epsfig}

\begin{document}

\def\be{\begin{equation}}
\def\ee{\end{equation}}
\def\rd{{\rm d}}
\newcommand{\nn}{\nonumber}
\newcommand{\ra}{\rightarrow}

% Added by PCT--start---------

\newcommand{\ba}{\begin{eqnarray}}
\newcommand{\ea}{\end{eqnarray}}
\newcommand{\Eq}[1]{Eq.~(\ref{#1})}
\newcommand{\Eqs}[1]{Eqs.~(\ref{#1})}
\newcommand{\rf}[1]{(\ref{#1})}
\newcommand{\Ref}[1]{Ref.~\cite{#1}}
\newcommand{\Refs}[1]{Refs.~\cite{#1}}
\newcommand{\Sec}[1]{Sec.~\ref{#1}}
\newcommand{\Secs}[1]{Secs.~\ref{#1}}
\newcommand{\Fig}[1]{Fig.~{\ref{#1}}}
\newcommand{\Figs}[1]{Figs.~{\ref{#1}}}
\newcommand{\Table}[1]{Table~{\ref{#1}}}

\title{Earth Matter Effects in Detection of Supernova Neutrinos\\}
\author{X.-H. Guo$^1\footnote{xhguo@bnu.edu.cn}$ and
  Bing-Lin Young$^{2,3}$\footnote{young@iastate.edu}}

\affiliation{$^1$ Institute of Low Energy Nuclear Physics, Beijing
Normal University, Beijing 100875, China}

\affiliation{$^2$ Department of Physics and Astronomy, Iowa State
University, Ames, Iowa 5011, USA}

\affiliation{$^3$ Institute of Theoretical Physics, Chinese Academy of Sciences, Beijing 100080, China}

\vspace{1cm}

\begin{abstract}
We calculated the matter effect, including both the Earth and
supernova, on the detection of neutrinos from type II supernovae
at the proposed Daya Bay reactor neutrino experiment. It is found
that apart from the dependence on the flip probability $P_H$
inside the supernova and the mass hierarchy of neutrinos, the
amount of the Earth matter effect depends on the direction of the
incoming supernova neutrinos, and reaches the biggest value when
the incident angle of neutrinos is around 93$^\circ$. In the
reaction channel $\bar{\nu}_e + p \ra e^+ + n$ the Earth matter
effect can be as big as about 12\%. For other detection processes
the amount of the Earth matter effect is a few per cent.
\end{abstract}

\maketitle

\section{\label{sec:intro} Introduction}

One of the most spectacular cosmic events is a supernova (SN)
explosion which ends the existence of a giant star. There are two
types of such explosions: types I and II. In the case of type I
supernovae the explosion gives out gigantic firework display,
while in the type II supernovae intensive neutrinos are emitted
for a very short period of time followed by an intensive
electromagnetic radiation.  Supernovae are a very good natural
laboratory to study various fundamental issues of physics and
astrophysics.

The type II SN explosion is caused by the mechanism of core
collapse. The total energy release is approximately the
gravitational binding energy of the core. It generates intensive
neutrinos which take away about 99\% of the total energy. The
explosion itself consumes about 1\% of the total energy. The
neutrino emission happens several hours prior to the brightening
of the SN which only consumes 0.1\% of the total energy. The
observation of the SN neutrinos can serve as an early warning for
the optical emission of a type II SN.

The first observations of the type II SN were made by Kamiokande
II , IMB, and Baksan in 1987, with totally 24 neutrino events
observed in a short period of about 13 seconds \cite{nuexp}.
Although this number of events is too small for a quantitative
study of neutrinos from supernova explosion, they are definitely
very valuable for providing the first signal of cosmic neutrinos
and initiating astrophysics and particle physics studies of
supernova neutrinos.

The Daya Bay underground neutrino laboratory has been proposed and
R\&D is now being carried out. It is located
%at the east side of the
%Dapeng peninsula, on the west coast of Daya Bay
in south China close to Hong Kong and provides an excellent site
for measuring the neutrino mixing angle $\theta_{13}$ by using 4
reactor cores in two clusters, each core with a thermal power of
2.9 GW and therefore 11.6 GW total power. Furthermore, two more
cores with an additional thermal power of 5.8 GW are expected to
be online in 2011. Since the site is close to a group of hills,
the cosmic ray background to the neutrino signals is vastly
reduced. While the main purpose of the Daya Bay experiment is to
measure a neutrino mixing angle, it can be used to detect possible
neutrino events from a SN explosion and serves as a part of the
Supernovae Early Warning System (SNEWS) \cite{SNEWS}. Hence the
theoretical prediction for the detection of SN neutrinos at Daya
Bay is very desirable.

In Ref. \cite{cadonati} the authors calculated the expected
neutrino events from a type II SN explosion in Borexino. For a
typical SN at a distance of 10 kpc a burst of around 110 events
would appear in the detector of Borexino. In this calculation,
matter effects are ignored.  Later Whisnant and Young \cite{young}
considered the detection of type II SN neutrinos in the Daya Bay
experiment with the matter effect inside the supernova being
included. This leads to the modified results which depend on the
flip probability of neutrinos while passing through resonance
regions inside a supernova.

In the realistic case, the neutrinos produced from a SN explosion
will likely go through some portion of the Earth before reaching
the detector. Therefore, in order to give a more accurate
prediction for the number of signal events, the Earth matter
effect should also be taken into account. The purpose of the
present paper is to calculate the Earth matter effect and
therefore takes into account all matter effects, both the Earth
and Supernova, on the detection of type II SN neutrinos in the
Daya Bay experiment.

\section{\label{sec:formalism} The formalism}

During a SN explosion, neutrinos are produced in two bursts. In
the first burst which lasts for only a few milliseconds, electron
neutrinos are generated via the inverse $\beta$-decay process $e^-
+ p \ra \nu_e +n$ which leads to a neutron rich star. In the
second burst which is longer ({$\cal O$}(10) seconds) neutrinos of
all flavors ($\nu_\alpha$ and $\bar{\nu}_\alpha$ with $\alpha$
being $e, \mu, \tau$) are produced via the $e^+ e^-$ annihilation
$e^+ + e^- \ra \nu_\alpha + \bar{\nu}_{\alpha}$, the electron
neutrino antineutrino annihilation $\nu_e + \bar{\nu}_{e} \ra
\nu_\alpha + \bar{\nu}_{\alpha}$, and nucleon-nucleon
bremsstrahlung $ N + N \ra N + N + \nu_\alpha +
\bar{\nu}_{\alpha}$ processes. While these neutrinos generated in
the SN explosion travel out of the SN, they interact with the high
density core of the SN. Neutrinos $\nu_x (\nu_x=\nu_\mu, \nu_\tau,
\bar{\nu}_\mu, \bar{\nu}_\tau)$ decouple first from deep inside
the core due to their lesser interactions as they only interact
via the neutral current. Then $\bar{\nu}_e$ decouples. The last
one to decouple is $\nu_e$ due to its interaction with the neutron
which has a higher density than the proton. The typical
temperatures of these neutrinos are in the following ranges
\cite{dighe}:
 \be T_{\nu_e}=3 - 4~{\rm MeV}, \;\; \;
 T_{\bar{\nu}_e}= 5 - 6~{\rm MeV}, \;\; \;T_{\nu_x}=7 - 9~{\rm
 MeV}. \label{2a} \ee

For the neutrino of flavor $\alpha$, the time-integrated neutrino
energy spectrum can be described by the Fermi-Dirac distribution,
%\cite{????},
 \be F_\alpha^{(0)}(E)=\frac{N_\alpha^{(0)}}{F_2
 T_\alpha^3}\frac{E^2}{exp(E/T_\alpha)+1}, \label{2b} \ee
where $E$ is the energy of the neutrino and $T_\alpha$ the
temperature as given in Eq.(\ref{2a}), $N_\alpha^{(0)}$ is the
total number of the neutrino of flavor $\alpha$, and $F_j$, where
$j$ is an integer, is defined by
 \be F_j =\int_0^\infty \frac{x^j}{exp(x)+1}dx.
 \label{2c} \ee
The average neutrino energy is then obtained from the distribution
function Eq. (\ref{2b})
 \be \langle E_\alpha^{(0)} \rangle =
 \frac{F_3}{F_2} T_\alpha. \label{2d} \ee
The total number of the
neutrino $\nu_\alpha$ is \be
N_\alpha^{(0)}=\frac{L_\alpha^{(0)}}{\langle E_\alpha^{(0)}
\rangle},\label{2e} \ee where $L_\alpha^{(0)}$ is the luminosity
of $\nu_\alpha$. Then the energy spectrum function can be
rewritten as \be F_\alpha^{(0)}(E)=\frac{L_\alpha^{(0)}}{F_3
T_\alpha^4}\frac{E^2}{exp(E/T_\alpha)+1}. \label{2f} \ee In the
simplest argument the luminosity of each flavor is the same
\cite{totani, langanke}. Hence \be
L_\alpha^{(0)}=\frac{0.99}{6}E_{SN}^{(0)}, \label{2g} \ee where
$E_{SN}^{(0)}$ is the total energy released during the SN
explosion.

When the SN neutrinos of each flavor are produced they are approximately
the effective mass eigenstates due to the extremely high matter density
environment.  While they propagate outward to the surface of the SN
they could experience level crossing (neutrinos jump from one mass
eigenstate to another) in the MSW resonance regions \cite{msw}.
There are two MSW resonance regions which are determined by the
two pairs of parameters $(\Delta m_{31}^2, {\rm sin}^2
2\theta_{13})$ which is referred to as the high resonance region
and $(\Delta m_{21}^2, {\rm sin}^2 2\theta_{12})$
%which is called
the low resonance region, where $\Delta{m}^2_{\rm kj}=m^2_{\rm
k}-m^2_{\rm j}$, i.e., the mass square difference of the k and j
neutrinos, and $\theta_{\rm jk}$ the corresponding mixing
angle. Let us denote the probability that the neutrinos jump from
one mass eigenstate to another at the high (low) resonance layer
by $P_H (P_L)$. The large mixing angle (LMA) solution of the solar
neutrino constrains the flip probability at the low resonance
region to be zero ($P_L=0$) \cite{dighe}. $P_H$ is a function of
the mixing angle ${\rm sin}\theta_{13}$ and $\Delta m_{31}^2$.
Since the value of the mixing angle ${\rm sin}^2 2\theta_{13}$ is
still not determined, in numerical calculation we let $P_H$ vary
between 0 and 1. The level crossing diagrams are different for
normal and inverted mass hierarchies \cite{dighe}. This leads to
different forms for neutrino flux of mass engenstates at the
surface of SN.

When the neutrinos arrive at the Earth after travelling through
the cosmic distance all the oscillation factors are averaged out
and there is no coherence between different mass eigenstates.
Hence the neutrino fluxes reaching the Earth are the following:
For the normal hierarchy ($\Delta m_{31}^2 > 0$),
\begin{eqnarray}
F_{\nu_e}^{(N)}&=&F_{\nu_x}^{(0)}+(|U_{e2}|^2P_H+|U_{e3}|^2(1-P_H))(F_{\nu_e}^{(0)}-F_{\nu_x}^{(0)}),\nn \\
F_{\bar{\nu}_e}^{(N)}&=&F_{\nu_x}^{(0)}+|U_{e1}|^2(F_{\bar{\nu}_e}^{(0)}-F_{\nu_x}^{(0)}),\nn \\
2F_{\nu_x}^{(N)}&=&F_{\nu_e}^{(0)}+F_{\nu_x}^{(0)}-(|U_{e2}|^2P_H+|U_{e3}|^2(1-P_H))(F_{\nu_e}^{(0)}-F_{\nu_x}^{(0)}),\nn \\
2F_{\bar{\nu}_x}^{(N)}&=&F_{\bar{\nu}_e}^{(0)}+F_{\nu_x}^{(0)}-|U_{e1}|^2(F_{\bar{\nu}_e}^{(0)}-F_{\nu_x}^{(0)}),
\label{2h}
\end{eqnarray}
and for the inverted hierarchy ($\Delta m_{31}^2 < 0$),
\begin{eqnarray}
F_{\nu_e}^{(I)}&=&F_{\nu_x}^{(0)}+|U_{e2}|^2(F_{\nu_e}^{(0)}-F_{\nu_x}^{(0)}),\nn \\
F_{\bar{\nu}_e}^{(I)}&=&F_{\nu_x}^{(0)}+(|U_{e1}|^2P_H+|U_{e3}|^2(1-P_H))(F_{\bar{\nu}_e}^{(0)}-F_{\nu_x}^{(0)}),\nn \\
2F_{\nu_x}^{(I)}&=&F_{\nu_e}^{(0)}+F_{\nu_x}^{(0)}-|U_{e2}|^2(F_{\nu_e}^{(0)}-F_{\nu_x}^{(0)}),\nn \\
2F_{\bar{\nu}_x}^{(I)}&=&F_{\bar{\nu}_e}^{(0)}+F_{\nu_x}^{(0)}-(|U_{e1}|^2P_H+|U_{e3}|^2(1-P_H))(F_{\bar{\nu}_e}^{(0)}-F_{\nu_x}^{(0)}),
\label{2i} \end{eqnarray}
where $U_{ei} (i=1,2,3)$ are the elements of the neutrino mixing
matrix.

Due to the smallness of the mixing angle $\theta_{13}$, we ignore
the terms proportional to ${\rm sin}\theta_{13}$, then Eqs.
(\ref{2h}) and (\ref{2i})
%become
can be simplified as follows: For the normal hierarchy,
\begin{eqnarray}
F_{\nu_e}^{(N)}&=&P_H {\rm sin}^2\theta_{12}F_{\nu_e}^{(0)}+(1-P_H {\rm sin}^2\theta_{12})F_{\nu_x}^{(0)},\nn \\
F_{\bar{\nu}_e}^{(N)}&=&{\rm cos}^2\theta_{12}F_{\bar{\nu}_e}^{(0)}+{\rm sin}^2\theta_{12}F_{\nu_x}^{(0)},\nn \\
2F_{\nu_x}^{(N)}&=&(1-P_H {\rm sin}^2\theta_{12})F_{\nu_e}^{(0)}+(1+P_H {\rm sin}^2\theta_{12})F_{\nu_x}^{(0)},\nn \\
2F_{\bar{\nu}_x}^{(N)}&=&{\rm
sin}^2\theta_{12}F_{\bar{\nu}_e}^{(0)}+(1+{\rm
cos}^2\theta_{12})F_{\nu_x}^{(0)}, \label{2j}
\end{eqnarray}
and for the inverted hierarchy ($\Delta m_{31}^2 < 0$),
\begin{eqnarray}
F_{\nu_e}^{(I)}&=&{\rm sin}^2\theta_{12}F_{\nu_e}^{(0)}+{\rm cos}^2\theta_{12}F_{\nu_x}^{(0)},\nn \\
F_{\bar{\nu}_e}^{(I)}&=&P_H{\rm cos}^2\theta_{12}F_{\bar{\nu}_e}^{(0)}+(1-P_H{\rm cos}^2\theta_{12})F_{\nu_x}^{(0)},\nn \\
2F_{\nu_x}^{(I)}&=&{\rm cos}^2\theta_{12}F_{\nu_e}^{(0)}+(1+{\rm sin}^2\theta_{12})F_{\nu_x}^{(0)},\nn \\
2F_{\bar{\nu}_x}^{(I)}&=&(1-P_H{\rm
cos}^2\theta_{12})F_{\bar{\nu}_e}^{(0)}+(1+P_H{\rm
cos}^2\theta_{12})F_{\nu_x}^{(0)}. \label{2k} \end{eqnarray}

If the SN matter effect is neglected, the neutrinos, which are
produced in the core of the SN and propagate through cosmic
distance to reach Earth, are subject to the vacuum oscillation only.
In this case when the Earth matter effect is also neglected the
neutrino fluxes at the detector are given by:
\begin{eqnarray}
F_{\nu_e}^{(V)}&=&(1-\frac{1}{2}{\rm sin}^22\theta_{12})F_{\nu_e}^{(0)}
               +\frac{1}{2}{\rm sin}^22\theta_{12}F_{\nu_x}^{(0)},\nn \\
F_{\bar{\nu}_e}^{(V)}&=&(1-\frac{1}{2}{\rm sin}^22\theta_{12})
               F_{\bar{\nu}_e}^{(0)}+\frac{1}{2}{\rm sin}^22\theta_{12}
               F_{\nu_x}^{(0)},\nn \\
2F_{\nu_x}^{(V)}&=&\frac{1}{2}{\rm sin}^22\theta_{12}F_{\nu_e}^{(0)}
               +(2-\frac{1}{2}{\rm sin}^22\theta_{12})F_{\nu_x}^{(0)},\nn \\
2F_{\bar{\nu}_x}^{(V)}&=&\frac{1}{2}{\rm sin}^22\theta_{12}
               F_{\bar{\nu}_e}^{(0)}+(2-\frac{1}{2}{\rm sin}^22
               \theta_{12})F_{\nu_x}^{(0)}.
 \label{2vac}
 \end{eqnarray}
\begin{figure}[htb]
\begin{center}
{\epsfig{file=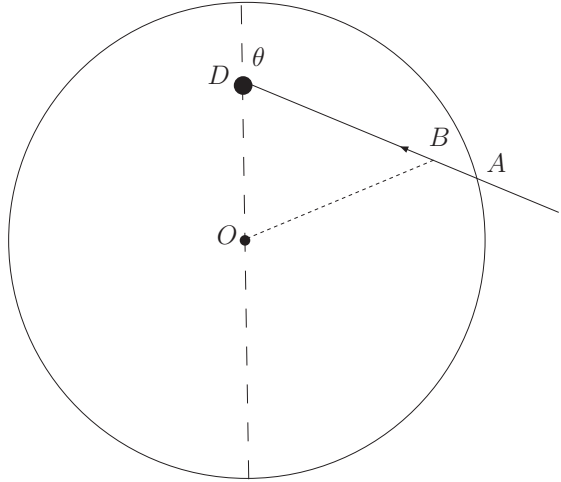, height=7.25cm, angle=-0}}
\caption{Illustration of the path of the SN neutrinos reaching
the detector in the Earth. $D$ is the location of the detector, $\theta$ is the
incident angle of neutrinos, and $O$ is the centre of the Earth.}
\label{fig1}
\end{center}
\end{figure}

In Eqs. (\ref{2j}) and (\ref{2k}) the Earth matter effect is
neglected. In reality, neutrinos from an SN may likely travel
through a significant portion of the Earth before reaching the
detector and are therefore subject to the matter effect. The Earth
matter effect can be parameterized in terms of the distance
between the neutrino and the Earth center and the total distance
travelled in the Earth before reaching the detector. The distance
that the neutrino travels in the Earth depends on their incident
orientation. Suppose a neutrino reaches the detector with the
incident angle $\theta$ (see Fig. 1), then the distance the
neutrino travels through the Earth is
 \be L=(-R+h){\rm cos} \theta +\sqrt{R^2-(R-h)^2 {\rm sin}^2\theta},
 \label{2l} \ee
where $h$ is the depth of the detector in and $R$ the radius of
the Earth.
%It can be seen from
%Eq. (\ref{2l}) that when the neutrinos come from the top and the
%bottom  of the detector corresponding to $\theta=0$ and
%$\theta=\pi$ respectively, the distance they travel before
%reaching the detector is $h$ and $2R-h$, respectively.
The distance of the neutrino to the Earth center can be obtained
as follows:  Let $x$ be the distance that the neutrino travels
into the Earth, which is shown as the segment $AB$ in Fig. 1. The
distance of the neutrino to the center of the Earth $\tilde{x}$ is
given by
 \be \tilde{x}=\sqrt{(R-h)^2+(L-x)^2+2(R-h)(L-x){\rm cos}
 \theta}. \label{2m} \ee

The Earth matter effects on the detection of SN neutrinos have been discussed extensively
in literature \cite{dighe, ioa1, ioa2, lunardini, dighe2, dighe3, lunardini2, mirizzi}. As
mentioned before, after travelling the cosmic distance to reach
the Earth, the arriving neutrinos are definite mass eigenstates
which, then, oscillate in going through the Earth matter. Let
$P_{ie}$ be the probability that a neutrino mass eigenstate
$\nu_i$ enters the surface of the Earth and arrives at the
detector as an electron neutrino $\nu_e$. Then the flux of $\nu_e$
at the detector, denoted as $F_{\nu_e}^D$, can be written as %
 \be F_{\nu_e}^D=\sum_i P_{ie}F_i, \label{2n} \ee %
where $F_i$ is the flux of $\nu_i$ at the Earth surface, in either
the normal or inverted hierarchy. $P_{ie}$ obey the unitary
condition $\sum_i P_{ie}=1$. In the case where the Earth matter
effect is ignored, $P_{ie}=|U_{ei}|^2$, then one recovers the
expressions of Eqs. (\ref{2h})-(\ref{2k}). After some
straightforward derivations with the terms proportional to ${\rm
sin} \theta_{13}$ being ignored, one obtains, in the case of the
normal hierarchy,
 \begin{eqnarray}
F_{\nu_e}^{D(N)}&=&P_{2e} P_H F_{\nu_e}^{(0)}+(1-P_{2e}P_H)F_{\nu_x}^{(0)},\nn \\
F_{\bar{\nu}_e}^{D(N)}&=&(1-P_{2e})F_{\bar{\nu}_e}^{(0)}+P_{2e}F_{\nu_x}^{(0)},\nn \\
2F_{\nu_x}^{D(N)}&=&(1-P_{2e}P_H)F_{\nu_e}^{(0)}+(1+P_{2e}P_H)F_{\nu_x}^{(0)},\nn \\
2F_{\bar{\nu}_x}^{D(N)}&=&P_{2e}F_{\bar{\nu}_e}^{(0)}+(2-P_{2e})F_{\nu_x}^{(0)},
 \label{2o}
 \end{eqnarray}
 and in the case of the inverted hierarchy,
\begin{eqnarray}
F_{\nu_e}^{D(I)}&=&P_{2e} F_{\nu_e}^{(0)}+(1-P_{2e})F_{\nu_x}^{(0)},\nn \\
F_{\bar{\nu}_e}^{D(I)}&=&P_H (1-P_{2e})F_{\bar{\nu}_e}^{(0)}+(1+P_{2e}P_H - P_H) F_{\nu_x}^{(0)},\nn \\
2F_{\nu_x}^{D(I)}&=&(1-P_{2e})F_{\nu_e}^{(0)}+(1+P_{2e})F_{\nu_x}^{(0)},\nn \\
2F_{\bar{\nu}_x}^{D(I)}&=&(1+P_{2e}P_H-P_H)F_{\bar{\nu}_e}^{(0)}+(1+P_H-P_{2e}P_H)F_{\nu_x}^{(0)}.
\label{2o1} \end{eqnarray}

 The probability $P_{ie}\;(i=1,2,3)$ has been calculated in Refs.
\cite{ioa1} and \cite{ioa2}. In the Earth the potential which an
electron neutrino experiences is
 \be V(x)=\sqrt{2}G_F N_e(x), \label{2p} \ee
where $G_F$ is the Fermi constant and $N_e (x)$ is the Earth
electron number density. Let us introduce a small parameter
 \be \epsilon(x)=\frac{2EV(x)}{\Delta m_{21}^2}. \label{2q} \ee
%where $E$ is the neutrino energy and $\Delta m_{21}^2$ is the mass
%square difference between the second and the first mass
%eigenstates of the neutrino, $\Delta m_{21}^2 =m_2^2-m_1^2$.
For a typical neutrino energy $E=10$ MeV, $\epsilon$ is less than
0.13. Neglecting contributions of $O(\epsilon^2)$, $P_{2e}$ can be
expressed as the following \cite{ioa1}:
 \be
P_{2e}={\rm sin}^2 \theta_{12} +\frac{1}{2}{\rm sin}^22
\theta_{12}\int_{x_0}^{x_f} dx V(x) {\rm sin} \phi_{x\ra x_f}^m,
 \label{2r} \ee
where $\theta_{12}\approx 33^\circ$ \cite{mixingangle12} and
$\phi_{a\ra b}^m$ is defined as
 \be \phi_{a\ra b}^m \equiv \int_a^b dx \Delta_m (x),
 \label{2s} \ee
where
 \be \Delta_m (x)\equiv \frac{\Delta m_{21}^2}{2E}\sqrt{(\cos
 2\theta _{12}-\epsilon(x))^2 + \sin^2 2\theta_{12}}.
 \label{2t} \ee

Let the matter density inside the Earth be $\rho (x)$
%where $x$ is the distance to the center of the Earth,
then the nucleon number density is $2\rho(x)/(m_p+m_n)$,
% ($m_p$ and $m_n$ are the masses of the proton and the neutron,
% respectively)
which, for matters of equal number of protons and neutrons, leads
to the following density of electrons inside the Earth:
 \be N_e (x)=\rho(x)/(m_p+m_n). \label{2u} \ee
In our calculations of the Earth matter effect we choose a simple
model for $\rho (x)$ which is called the mantle-core-mantle
density profile given in \cite{stacey} and \cite{freund}. In this
model a step function is used to describe the Earth matter effect:
$\rho=12 g/cm^3$ for the core, and $\rho=5 {\rm g}/{\rm cm}^3$ for
the mantle. The core radius and the thickness of the mantle are
each half of the Earth radius.

In the next section we will use the formulas given in this section
to calculate numerically the neutrino fluxes at the detector,
taking into account the Earth and supernova matter effects, and
then apply the results to various physical channels through which
neutrinos can be detected.

\section{\label{sec:numerical results} Numerical Results}

In the following we calculate the predicted numbers of events that
can be observed through various reaction channels at the Daya Bay
experiment. This will be done by integrating, over the neutrino
energy $E_\nu$, the product of the target number $N_T$, the cross
section of each channel, and the neutrino flux function
$F_\alpha^D (E_\nu)/4\pi D^2$ given in Eqs. (\ref{2o}) and
(\ref{2o1}) \cite{young}:
 \be N(i) =N_T \int dE_\nu \sigma (i) (E_\nu)
 \frac{1}{4\pi D^2}F_\alpha^D (E_\nu), \label{3a} \ee
where $D$ is the distance between the SN and the Earth and the index
$i$ represents different channels through which the SN neutrinos
are observed.

The target material includes electrons, protons and carbon nuclei
since the chemical composition of the expected liquid
scintillator, mesitylene or pseudocumene, is $C_9 H_{12}$ which is
the same as that of Borexino \cite{cadonati}. The cross section
for each reaction channel can be found in Ref. \cite{cadonati}.
For neutrino-electron elastic scatterings via charged and neutral currents, 
$\nu_\alpha (\bar{\nu}_\alpha) +e^- \ra \nu_\alpha
(\bar{\nu}_\alpha) +e^-$, the cross sections are:
\begin{eqnarray}
\sigma (\nu_e e) &=&9.2 \times 10^{-45} E_\nu ({\rm MeV}) \;{\rm cm}^2,\nn \\
\sigma (\bar{\nu}_e e) &=&3.83 \times 10^{-45} E_\nu ({\rm MeV}) \;{\rm cm}^2,\nn \\
\sigma (\nu_{\mu, \tau} e) &=&1.57 \times 10^{-45} E_\nu ({\rm MeV}) \;{\rm cm}^2,\nn \\
\sigma (\bar{\nu}_{\mu, \tau} e) &=&1.29 \times 10^{-45} E_\nu
({\rm MeV}) \;{\rm cm}^2. \label{3b} \end{eqnarray}

The inverse beta decay $\bar{\nu}_e + p \ra e^+ +n$ has the most
number of events due to its large cross section and low threshold
of the neutrino energy.
\begin{eqnarray}
\sigma (\bar{\nu}_e p) &=&9.5 \times 10^{-44} (E_\nu({\rm
MeV})-1.29)^2 \;{\rm cm}^2,\nn \\
E_{\rm th}&=&1.80~{\rm MeV}. \label{3c} \end{eqnarray}

For the neutrinos and $^{12}C$ system, there are two charged-current 
and three neutral-current reactions. Their effective cross
sections are obtained by scaling the experimentally measured energy values
from muon decay at rest to the energy scale for supernova neutrinos. In this
way, one obtains the following effective cross section \cite{cadonati}:
\begin{description}
\item charged-current capture of $\bar{\nu}_e$:
 $\bar{\nu}_e+ ^{12}C \ra  ^{12}B + e^+, \;%\;
 ^{12}B \ra ^{12}C + e^- + \bar{\nu}_e$,
 \be \langle\sigma (^{12}C(\bar{\nu}_e, e^+) ^{12}B)\rangle =1.87
 \times 10^{-42} \;cm ^2; \label{3d} \ee
\item charged-current capture of $\nu_e$:
 $\nu_e + ^{12}C \ra  ^{12}N + e^-, \;%\;
 ^{12}N \ra ^{12}C + e^+ + \nu_e$,
 \be \langle\sigma (^{12}C(\nu_e, e^-) ^{12}N)\rangle =1.85 \times
 10^{-43} \;cm ^2; \label{3e} \ee
\item neutral-current inelastic scattering of $\nu_\alpha$ or
$\bar{\nu}_\alpha$:
 $\nu + ^{12}C \ra  ^{12}C^* + \nu', \;%\;
 ^{12}C^* \ra ^{12}C + \gamma$,
 \begin{eqnarray}
 %\be
 \langle\sigma (\nu_e ~^{12}C)\rangle &=& 1.33 \times 10^{-43} \;cm
^2; \label{3f} \\ %\ee
 %\be
 \langle\sigma (\bar{\nu}_e ~^{12}C)\rangle
 & = & 6.88 \times 10^{-43} \;cm ^2; \label{3g} \\ %\ee
 %\be
 \langle\sigma (\nu_x(\bar{\nu}_x) ~^{12}C)\rangle &=& 3.73 \times 10^{-42} \;cm
^2,\;\;x=\mu,\tau. \label{3h} %\ee
\end{eqnarray}
\end{description}
These reactions will be referred to collectively as
neutrino-carbon scatterings.

There will be several detectors located in the near and far sites.
We take the total detector mass to be 300 tons which is the same
as that of Borexino \cite{cadonati}. Then the total numbers of
target electrons, protons, and $^{12}C$ are
 $$N_T^{(e)}=9.94 \times 10^{31},\;\; N_T^{(p)}=1.82
 \times 10^{31},\;\;N_T^{(C)}=1.36 \times 10^{31}.$$
In the numerical calculations we take the total energy release of
SN as
$$E_{SN}^{(0)}=3 \times 10^{53}~{\rm erg} =1.97 \times 10^{59}~{\rm MeV},$$
and the distance of SN as
$$D=10~{\rm kpc} =3.09 \times 10^{22}~{\rm cm}.$$
Other parameters are taken as follows:
 $$h=0.4~{\rm km}, \;\; R=6400~{\rm km},\;\;
 \Delta m_{21}^2=7.1 \times 10^{-5}~{\rm eV}^2,\;\;
 \theta_{12}=32.5^\circ.$$

The numerical results for the event numbers for various reaction
channels are obtained from Eqs. (\ref{3a})-(\ref{3h}), (\ref{2o})
and (\ref{2o1}). The results are shown in Figs. 2, 3, and 4. It
can be seen from these figures that the Earth matter effect
depends on the incident angle of the neutrino, the mass hierarchy,
and the flip probability $P_H$. When the incident angle of the SN
neutrino is smaller than a value $\theta_0$, where $\theta_0 \sim
90^\circ$,
%the inverse beta decay, neutrino-electron scatterings, and
%neutrino-carbon scatterings,
the Earth matter effect can be ignored for all the relevant
reactions that concern us here.

The inverse beta decay $\bar{\nu}_e + p \ra e^+ +n$ has the
largest Earth matter effect among all the channels. The maximum
Earth matter effect for the inverse beta decay appears at around
$\theta \sim 93.6^\circ$ in the cases of $P_H=0, 1$ (normal
hierarchy) and $P_H=1$ (inverted hierarchy) and the effect is
12.0\%. When the incident angle become larger than about
$106^\circ$, the Earth matter effect becomes about 6.7\% and
almost independent of the incident angle.

For the neutrino-electron and neutrino-carbon scattering channels
the behavior is quite similar to that of the inverse beta decay.
For neutrino-electron elastic scattering, the maximum Earth matter
effect appears at $\theta \sim 92.7^\circ$ and the amount could be
as large as 3.75\% in the case of $P_H=1$ for both normal and
inverted hierarchies, 3.11\% in the case of $P_H=0$ for the
inverted hierarchy, and 0.66\% in the case of $P_H=0$ for the
normal hierarchy.  When the incident angle become larger
than about $100^\circ$, the Earth matter effect is independent of
the incident angle and the amount is about 2.16\% for $P_H=1$
(normal and inverted hierarchies), 1.76\% for $P_H=0$ (inverted
hierarchy), and 0.38\% for $P_H=0$ (normal hierarchy),
respectively.

It should be noted that there are complications in dealing with
reactions of neutrinos with $^{12}C$. The effective cross sections
in Eqs. (\ref{3d}-\ref{3h}) are given for supernova neutrinos
without oscillations. It is pointed out in Ref. \cite{cadonati}
that the effective cross sections are affected when neutrino
oscillations are taken into account. For instance, the oscillation
of higher energy $\nu_\mu$ into $\nu_e$ results in an increased
event rate since the expected $\nu_e$ energies are just at or
below the charged-current reaction threshold. This leads to an
increase by a factor of 35 for the cross section $^{12} C(\nu_e
e^-) ^{12}N$ if one average over a $\nu_e$ distribution with $T=8$
MeV rather than 3.5 MeV.  After numerical calculations for  the
reactions of neutrinos with $^{12}C$, we find that the Earth
matter effects become the largest when $\theta \sim 91.8^\circ$
and the maximum amount is 4.00\% in the case of $P_H=1$ for both
normal and inverted hierarchies, 1.65\% in the case of $P_H=0$ for
the inverted hierarchy, and 2.00\% in the case of $P_H=0$ for the
normal hierarchy.  Again when the incident angle become larger
than about $100^\circ$, the Earth matter effect is almost independent of
the incident angle and the amount is about 2.87\% for $P_H=1$
(normal and inverted hierarchies), 1.21\% for $P_H=0$ (inverted
hierarchy), and 1.36\% for $P_H=0$ (normal hierarchy),
respectively.
%there are two peaks for the Earth matter effect, one at
%$\theta=90.9^\circ$ and the other $\theta=92.7^\circ$. However,
%the amount is smaller than 0.6\%. When $\theta$ is larger than
%about $100^\circ$, the Earth matter effect becomes negligible.

The above results can be understood from the fact that the
oscillation behavior is determined by the phase factor
$\Delta{m}^2_{21}({\rm eV}^2)L({\rm m})/E({\rm MeV})$, where $L$
is the distance in the Earth that the neutrino travels.  Equation
(\ref{2l}) gives the distance the SN neutrinos travel in the
Earth. It can be seen from this equation that when $\theta <
\theta_0$  this distance is smaller than 10 km and hence the
amount of Earth matter effect is very small. In this case $P_{2e}$
in Eq. (\ref{2r}) and $F_{\nu_e}^D$ in Eq. (\ref{2n}) can be
replaced by their values in the vacuum $|U_{e2}|^2$ and
$F_{\nu_e}$, respectively. When $\theta$ is larger than $90^\circ$
this distance becomes bigger than 100 km, the Earth matter effect
becomes large and reaches a maximum for $\theta \sim 91^\circ -
94^\circ$. When $\theta$ is larger than $100^\circ$, the distance
that neutrinos travel in the Earth is greater than 2000 km, then
there could be many oscillations in $F_{\nu_e}^D$. This leads to
averaging the Earth matter effect which is smaller than the
maximum value.  As an example, the flux of the SN neutrino
$\bar{\nu}_e$ as a function of neutrino energy in various cases
are plotted in Fig. 5. It can be seen from Figs. 2 to 5 that the
event numbers detected at the detector depend on the incident
angle of neutrinos, the SN matter effects ($P_H$), and the mass
hierarchy of neutrinos. By measuring these event numbers we can
obtain information on the SN and the Earth matter effect, and the
neutrino mass hierarchy.

\vspace{2cm}
\begin{figure}[htb]
\begin{center}
 {\epsfig{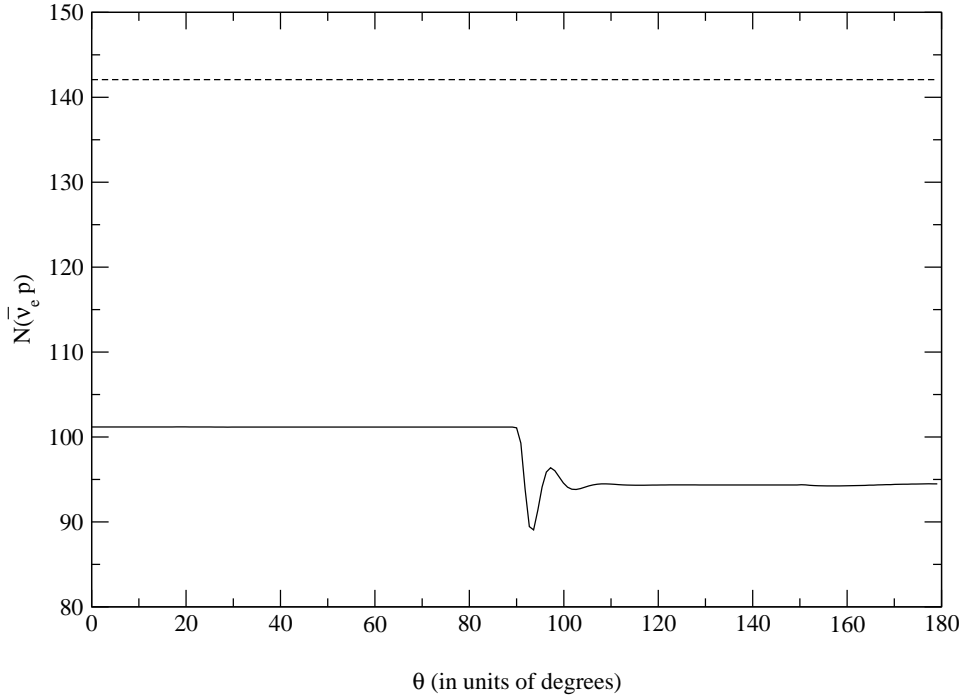}}
\caption{The event number of the reaction $\bar{\nu}_e + p \ra e^+
+n$ as a function of the incident angle of the SN neutrino. The
solid curve corresponds to $P_H=0, 1$ (normal hierarchy) and
$P_H=1$ (inverted hierarchy). The dashed curve corresponds to
$P_H=0$ (inverted hierarchy). } \label{fig2}
\end{center}
\end{figure}

\vspace{2cm}
\begin{figure}[htb]
\begin{center}
 {\epsfig{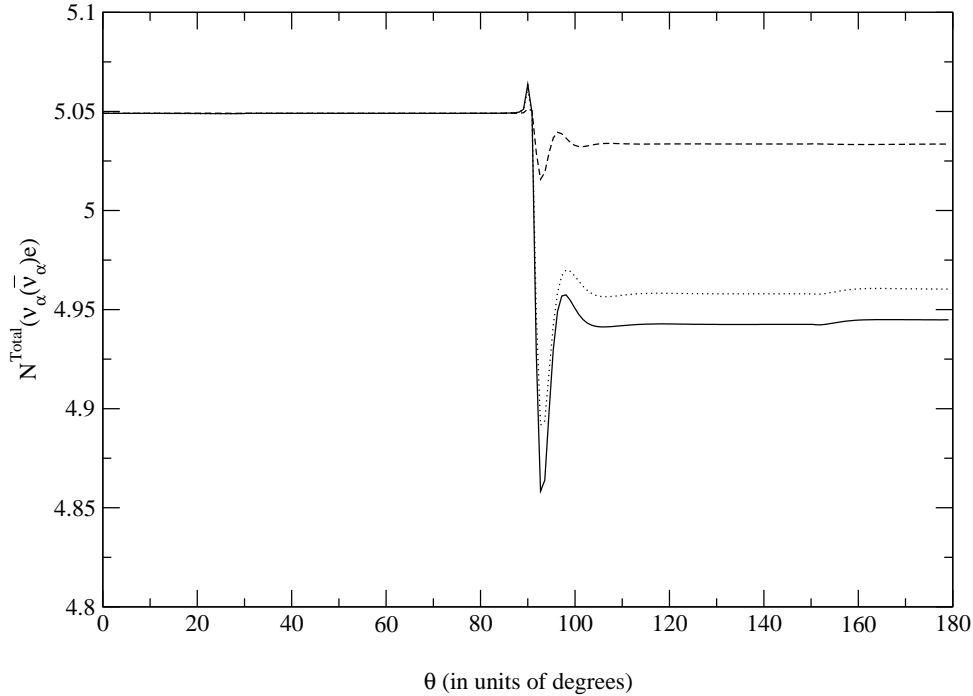}}
\caption{The total event number of neutrino-electron elastic
scattering via charged and neutral currents $\nu_\alpha
(\bar{\nu}_\alpha) +e^- \ra \nu_\alpha (\bar{\nu}_\alpha) +e^-$ as
a function of the incident angle of the SN neutrinos. The solid
curve corresponds to $P_H=1$ (normal and inverted hierarchies), the
dashed curve corresponds to $P_H=0$ (normal hierarchy), and the
dotted curve corresponds to $P_H=0$ (inverted hierarchy),
respectively. } \label{fig3}
\end{center}
\end{figure}

\begin{figure}[htb]
\begin{center}
 {\epsfig{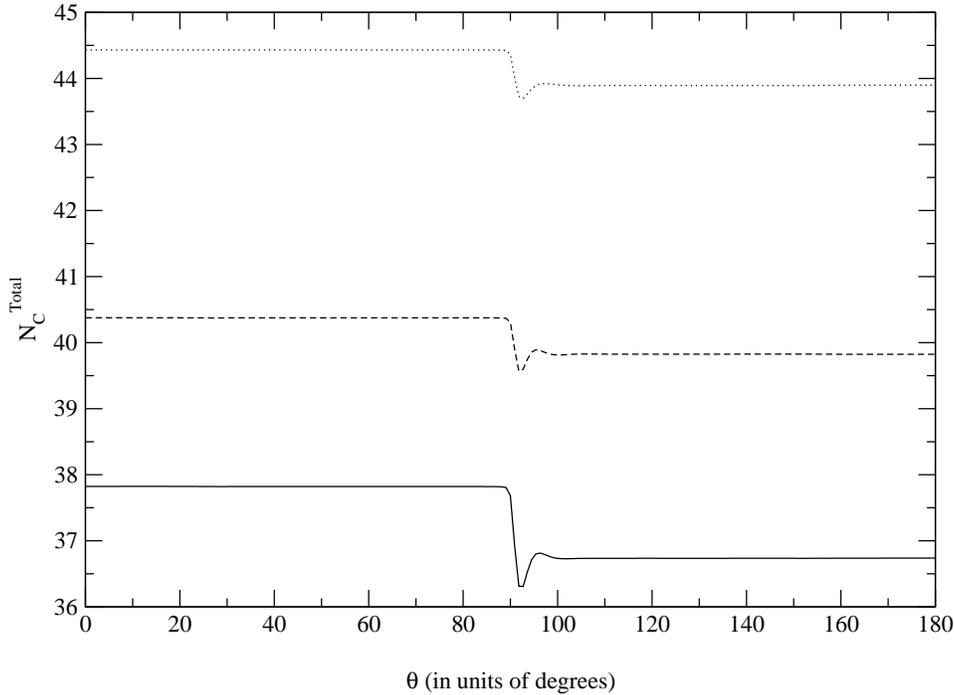}}
\caption{The total event number of the reactions of neutrinos with
$^{12}C$ as a function of the incident angle of the SN neutrinos.
The solid curve corresponds to $P_H=1$ (normal and inverted
hierarchies), the dashed curve corresponds to $P_H=0$ (normal
hierarchy), and the dotted curve corresponds to $P_H=0$ (inverted
hierarchy), respectively.} \label{fig4}
\end{center}
\end{figure}

\begin{figure}[htb]
\begin{center}
 {\epsfig{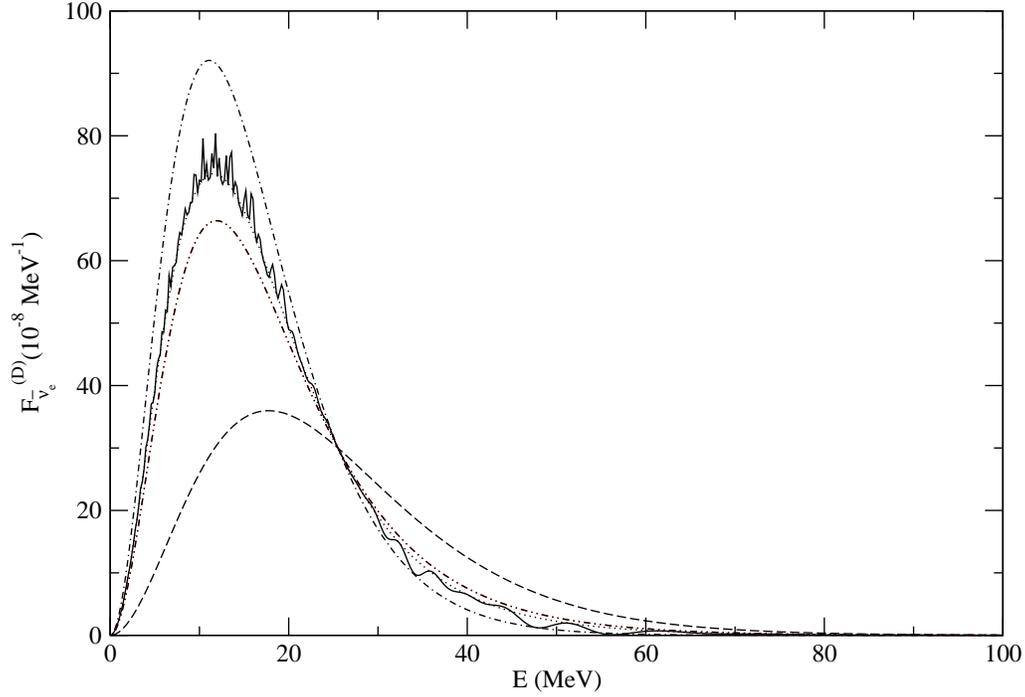}}
\caption{The flux of the SN $\bar{\nu}_e$ at the detector
multiplied by $10^{-8}$. The solid curve represents the cases
where supernova and Earth matter effects are included for
$\theta=180^\circ$, $P_H=0,1$ (normal hierarchy) or $P_H=0$
(inverted hierarchy). The dashed curve represents the case where
supernova and Earth matter effects are included for
$\theta=180^\circ$, $P_H=0$ (inverted hierarchy) or the case where
the supernova matter effect is included but the Earth matter
effect is not for $P_H=0$ (inverted hierarchy).  The dotted
curve represents the cases where the supernova matter effect is
included but the Earth matter effect is not for $P_H=0,1$ (normal
hierarchy) or $P_H=1$ (inverted hierarchy). The dot-dashed
(dot-dot-dashed) curve represents the
flux given by Eq. (\ref{2f}) (Eq. (\ref{2vac})) where %both the
%Earth matter effect and the SN matter effect are not included and
the neutrinos do not oscillate at all (oscillate as in vacuum).}
\label{fig5}
\end{center}
\end{figure}

\section{\label{sec:summary} Summary}

We have calculated matter effects, including both the supernova
and the Earth, on the detection of type II supernovae neutrinos at
the proposed Daya Bay experiment. It is found that the amount of
the matter effect depends on the neutrino incident angle, the
neutrino mass hierarchy, and the flip probability $P_H$. When the
incident angle of the SN neutrinos is smaller than a definite
value $\theta_0$, the Earth matter effect is negligible due to the
small distance that the neutrino traverses in the Earth. The Earth
matter effect becomes manifested and reachs a maximum value for
$\theta \sim 91^\circ - 94^\circ$. When $\theta$ is greater than
$100^\circ$, the Earth matter effect becomes insensitive to the
neutrino incident angle and the effect is smaller than the maximum
value. There is a window in $\theta$, $\theta\sim
90^\circ-100^\circ$, corresponding to the neutrino travelling
distance of 500 km to 2000 km, in which the neutrino event numbers
can vary significantly as a function of $\theta$. This phenomenon
can be understood by the fact that the Earth matter effect is
controlled by $\Delta{m}^2_{21}$ and the variation of the neutrino
number is significant for $\Delta{m}^2_{21} ({\rm eV}^2) L({\rm
m})/E_\nu({\rm MeV})\sim$ a few.

In the reaction channel $\bar{\nu}_e + p \ra e^+ + n$  the amount
of the Earth matter effect is the largest and the maximum amount
is about 12\%. For the neutrino-electron and neutrino-carbon
scattering channels, the amount of the Earth matter effect is a
few per cent at most. By measuring the event numbers in various
channels we can obtain information on the Earth matter effect, the
matter effect inside the SN, and the neutrino mass hierarchy.

\begin{acknowledgments}
We would like to thank members of the Daya Bay Collaboration for
helpful discussions. This work was supported in part by the Special Grants
for 'Jing Shi Scholar' of Beijing Normal University.

\end{acknowledgments}

\end{document}